\documentclass[prb,preprint,groupedaddress,preprintnumbers]{revtex4}

\usepackage[T1]{fontenc}

\usepackage{dcolumn}

\usepackage{color,graphics,graphicx,fancybox,fancyhdr}
\usepackage{latexsym,amsmath,amssymb,accents,bm}

\begin{document}
\bibliographystyle{apsrev}

\graphicspath{D:\DATI\Bc1_data\bc1_jap} \DeclareGraphicsExtensions{.eps,.ps}

\title{First critical field measurements of superconducting films by third harmonic analysis}

\author{G. Lamura$^{a}$, M. Aurino$^{b}$ and A. Andreone}
\affiliation{CNR-INFM Coherentia and Department of Physics,
University of Naples "Federico II", I-80125, Napoli, Italy.}

\author{J.-C. Vill\'egier}
\affiliation{CEA, Institut Nanosciences et Cryogenie/SPSMS, 17, rue des Martyrs, 38054 Grenoble, France.}

\date{\today}
\begin{abstract}

The temperature behaviour of the first critical field ($B_{C1}$) of superconducting thin film samples can be determined with high accuracy using an inductive and contactless method. Driving a sinusoidal current in a single coil placed in front of the sample, a non zero third harmonic voltage $V_{3}$ is induced in it when Abrikosov vortices enter the sample. Conditions to be satisfied for the quantitative evaluation of $B_{C1}$ using this technique are detailed. As validation test, different type II superconductors (Nb, NbN, MgB$_{2}$ and Y$_{1}$Ba$_{2}$Cu$_{3}$O$_{7-d}$ under the form of thin films) have been measured. The comparison between experimental results, data presented in literature and theoretical predictions is presented and discussed. 

\end{abstract}


\maketitle

\section{INTRODUCTION}

The first critical field, $B_{C1}$, is the field at which quanta of magnetic flux (Abrikosov vortices) start to penetrate in a type II superconductor, and the so called mixed state sets in. The most popular method to determine this field is the measurement of the first magnetization cycle M(H) by using, for example, vibrating sample magnetometers (VSM) or superconducting quantum interference devices (SQUID) \cite{POOLE}. The effective $H_{C1}$ is then determined as the field at which M(H) departs from linearity, looking at the slope ($dM/dH$)=-1 ~ \cite{Hc1SLOPE} or as the field at which the induction B is different from zero \cite{Hc1TH}. This method is very effective for medium-high field values whereas it lacks in precision for superconducting materials whose first critical field is as low as the remanent magnetic moment. In this case the measurement of the remanent magnetic moment itself has been proposed to determine $H_{C1}$ \cite{REMANENT}.

An interesting and alternative way to determine the first critical field is the study of high order harmonics in the spectrum of the electrodynamic response of a superconductor to an alternating magnetic field excitation. This method, suggested years ago \cite{SHAULOVI}, is based on the superposition of a small sinusoidal field on a collinear steady field and has been also proposed to determine the irreversibility line ($T_{irr}$) \cite{SHAULOVI,BEEKII,BEEK}, below which magnetization becomes irreversibile as a result of flux pinning. However, the limits of validity of this technique to give a quantitative and reliable measurement of $B_{C1}$ has never been discussed in detail.

As a general feature, the generation of odd harmonics is predicted by the Bean  critical state model \cite{BEAN}. This model assumes that the supercurrent density circulating in the sample does not exceed a critical value that is independent from the applied magnetic field whereas it is a function of the Bean critical field $H^{*}$, beyond which \textit{all} the sample volume is penetrated by the magnetic field under the form of Abrikosov vortices. It is well known that this model does not take into account the existence of the lower critical field. In the  critical state the magnetization is irreversible, which means that losses are present and proportional to the area enclosed by the magnetization loop \cite{BEAN,TINKHAM}. In particular, the magnetization is proportional to the applied magnetic field whereas it does not depend explicitly on time. This renders the Bean model a static description of the mixed regime that completely neglects the flux creep contribution to losses \cite{BRANDT}. The hysteretic behavior of the magnetization in the critical state gives rise to non zero odd harmonics in the spectrum of the electrodynamic response of superconductors exposed to an ac magnetic field \cite{BEAN,TINKHAM}. Even harmonics can appear too, when a dc magnetic field is superposed to an ac magnetic field \cite{TINKHAM,CAMPBELL}. These features attracted a large interest in the scientific community for theoretical analysis and measurement of high order harmonics as a function of temperature, field and frequency to study the static \cite{SHAULOVI,DUBOIS,BEEKII,BEEK,SAMOILOV,SHATZ,WOLFUS,FARINON,DEAK,GIOACCHINO,PROZOROV,QIN,SAHOO} and dynamic properties of the flux line lattice\cite{POLI_I,POLI_II,POLI_III}. A useful review on the vortex regime studied by high order harmonics ac susceptibility has been presented in ref. \cite{GOMORY}.

The aim of this paper is to consider which conditions have to be satisfied for the use of the third harmonic analysis as a probe of the first critical field. As validation test, we will show experimental results on different type II superconductors, namely Nb, NbN, MgB$_{2}$ and Y$_{1}$Ba$_{2}$Cu$_{3}$O$_{7-d}$ (YBCO), in thin film form. The way $B_{C1}$ values extracted by using this method compare with theory and experimental results obtained by other techniques will be also discussed.

\section{THIRD HARMONIC ANALYSIS}

We consider the case of an ac magnetic field of amplitude \textit{h} and frequency \textit{f} generated by a far source applied to a superconducting infinite cylinder with the main axis parallel to the field direction (demagnetizing factor N=0). As \textit{h} equals $H_{C1}(T)$, vortices start to penetrate into the superconducting sample. Under these conditions a non linear power law $J \propto E^{n}$ in the current-voltage curve applies, where the exponent $n\rightarrow \infty$ in the Bean limit, $n>1$ in the flux creep regime, $n=1$ in the flux flow linear regime. As a consequence, odd harmonic components are produced in the spectrum of the sample response signal \cite{BRANDT,BADI}, \textit{when it enters into a region of field and temperature in the magnetic phase diagram delimited by $B_{C1}(T)$ and the irreversibility field $B_{irr}(T)$}. In the following, we will concentrate our attention to the temperature, field and frequency behavior of the third harmonic component $V_{3}$.
In the Bean limit, $V_{3}$ is predicted to be proportional to \cite{BEAN}:

\begin{eqnarray}\label{1}
V_{3} \propto h^{2}\cdot \frac{f}{J_{C}}.
\end{eqnarray}

By increasing the temperature in the superconducting state at fixed $h$ and \textit{f}, $V_{3}$ presents a maximum related to the Bean's critical field $H^{*}$ \cite{TINKHAM,GHATAK}. This maximum depends on the interplay between the amplitude of the applied field \textit{h} and $H^{*}$, whose temperature dependence follows the Bean critical current density ($H^{*}(T)\propto J_{C}$(T)). At low temperature and $h<H^{*}$, $V_{3}$ is determined by eq. 1, therefore it increases with temperature since $J_{C}$ decreases. As the temperature approaches the superconducting critical temperature $T_{C}$, the critical current density is so low that the condition $h>H^{*}$ is easily verified. In this regime $V_{3}\propto H^{*}$, thus it decreases with temperature as the critical current does. By summarizing, $V_{3}(T)$ is strictly equal to zero in the Meissner phase, assumes values different from zero with a bell-shape behavior in the mixed state below the irreversibility line, and it is again equal to zero in the flux flow and normal state regimes. The shape of the $V_{3}(T)$ peak is an important parameter to state the quality of the sample. If the zero temperature limit of the critical current density is low, the third harmonic peak is extremely broadened \cite{TINKHAM}. Moreover, multiple peaks can appear when regions with different values of $H^{*}$ are present, as it was shown experimentally in ref. \onlinecite{SHAULOVII}. 
We will refer to the left and right onset of the third harmonic curve as the temperatures at which $V_{3}$ starts to assume values different from zero for rising and lowering T respectively. To better clarify these important features, we have presented the third harmonic temperature behaviour respect to a generic superconducting phase diagram in figure 1.

The right onset of the third harmonic peak is mostly ascribed to the irreversibility temperature \cite{SHAULOVI,BEEKII,BEEK}. It is sensitive to the frequency \cite{SAMOILOV,BEEK,WOLFUS,GIOACCHINO,QIN,SAHOO} and the amplitude of  $h(t)$ \cite{SAMOILOV} and also to the dc magnetic field intensity \cite{BEEKII,WOLFUS,SHATZ,DUBOIS}. Since the irreversibility line is directly dependent on the pinning mechanism, its frequency dependence can assume different behaviours in different superconducting systems as it has been put in evidence for YBCO and BSCCO \cite{WOLFUS}.

Despite all the interest in the nonlinear electrodynamic properties of superconductors, the direct relation between the left onset of $V_{3}(T)$ and $H_{C1}$ has never been put in evidence properly. In order to justify this association, \textit{a discussion on the field and frequency dependence of $V_{3}$ left onset is essential}. As to the former, we observe from previous works that it decreases with increasing amplitude of the applied ac and dc magnetic field \cite{SHAULOVI,DUBOIS,BEEKII,BEEK,SAMOILOV,SHATZ,SAHOO}, as expected. As to the latter, we remember that $V_{3}(T)$ shows different frequency behaviours depending on the absence/presence of a dc magnetic field superposed to \textit{h(t)}. In the absence of dc field, and at constant $h(t)$ amplitude and low frequency ($f<1 kHz$), the $V_{3}(T)$ left onset is $f$-independent \cite{SAHOO} whereas the peak position shifts towards lower temperatures and its amplitude increases with increasing frequency. All these features have been experimentally verified up to 5KHz on YBCO thin films. For higher frequencies, $V_{3}(T)$ behaviour is predicted to be dependent on different sources of non linearity \cite{GIOACCHINO}. A further complication arises in the case of granular samples, where the measured first penetration field is the Josephson first critical field B$_{C1J}$, related to the nucleation of Josephson vortices and always lower than the intrinsic B$_{C1}$. B$_{C1J}$ is expected to be frequency independent up to the Josephson plasma frequency \cite{GUREVICH}, that is usually comparable or larger than 10 GHz \cite{HALBRITTER}. If a dc field is present, by increasing the frequency the left onset and the peak position shift towards higher temperatures and $V_{3}$ amplitude decreases \cite{PROZOROV,SHAULOVI,FARINON,BEEK,QIN}. This has been explained by taking into account flux creep effects \cite{QIN}.

For all these reasons, we focus our attention on the $V_{3}$ temperature behavior in the limit of low frequency ($f<1-5$ kHz seems to be a good compromise) and in zero dc applied field. Under these conditions, its measurement provides information on the left side onset of the third harmonic response, that can be identified as due to the first vortex penetration, and the associated value of the applied magnetic field can be taken as the first critical field at that temperature.

\section{EXPERIMENTAL}

The above discussion provides us a method to evaluate the first critical field in different type II superconductors under the form of thin films. To measure the third harmonic component of the sample response to an alternate magnetic field we used the experimental set up shown in figure 2a. A sinusoidal current, the reference signal of a Lock-in, is amplified and measured by using a high precision resistor. This signal is injected in a pancake coil (external diameter 4 mm) located at the end of a cryogenic insert. The sample is placed in front of the coil at a fixed distance ($h_{SC}$). In figure 2 b a cross section of the sample-coil configuration is represented. Details of the coil characteristics have been presented in ref. \cite{AURINO}. The output signal is filtered by using a low noise notch filter to reduce the first harmonic component and to avoid any saturation of the Lock-in for a better signal-to-noise ratio during third harmonic detection. 

In order to exactly evaluate the maximum total induction field at the sample surface, we used the model developed in ref. \onlinecite{AURINO} in which a superconducting infinite slab of thickness \textit{d} is placed at a distance $h_{SC}$ from a pancake coil. This model is valid only when the sample dimensions are larger than the external coil diameter. Instead of solving the problem in terms of the external magnetic field H generated by the pancake coil, it is easier to refer to the vector potential $\overrightarrow{A}$ since the current injected in the coil I$_{0}$ lies in a plane parallel to the sample surface. Therefore, as in a mirror image, the induced screening supercurrents will be localised in the sample plane. This consideration, and the assumption of a sample with in-plane isotropic conductivity $\sigma$, imply that only
the orthoradial component of the vector potential $\overrightarrow{A}$ expressed in cylindrical coordinates is different from zero. The total induction field components $B_{r}$, $B_{z}$ and $B_{\theta}$ on the sample surface have been exactly calculated \cite{AURINO}. $B_{\theta}$ is zero, $B_{z}$ is completely negligible, and the only component different from zero is $B_{r}$, since the total induction field is the sum of the applied ac magnetic field and the sample magnetization. The maximum amplitude of $B_{r}$ is:

\begin{eqnarray}\label{2}
B^{MAX}_{r}=\mu_{0} K I_{0}
\end{eqnarray}

where \textit{K} is the geometrical coil factor, that in our experimental configuration is set to $2 \cdot 10^{5}$ $m^{-1}$ ($h_{SC}$=100 $\mu$m).
Our sample-coil geometry gives access only to the first critical field in the direction orthogonal to the sample surface. Therefore, in the following we will refer to the measured B$_{C1}$ as B$^{\bot}_{C1}$.
The demagnetization factor is already taken into account in the extraction of the total induction field at the sample surface under the fundamental hypotesis of infinite slab \cite{AURINO}.

To validate this technique as a powerful and simple probe of the first critical field in type II superconducting planar samples, we have tested a number of selected thin films having low, medium and high critical temperature: \textit{i}) a 500 nm thick Nb film prepared by ultra-high vacuum cathodic arc deposition\cite{RUSSO} on Si wafers with T$_{C}$=9.2 K;  \textit{ii}) a 300 nm thick NbN (1111) textured film grown by DC magnetron sputtering \cite{NBN} on $(1 \bar{10} 2)-Al_{2}O_{3}$ with T$_{C}$=14.5 K; \textit{iii}) a c-axis oriented 400 nm MgB$_{2}$ epitaxial film grown on (0001)4H$-$SiC by HPCVD \cite{XI} with T$_{C}$=40.5 K; \textit{iv}) a 700 nm thick YBCO thin film produced by thermal evaporation (Theva GmbH) on (100) $LaAlO_{3}$ with T$_{C}$=87.9 K. The surface area of all these samples is 10x10 mm$^{2}$. In table I a summary of the main superconducting properties of the samples under test is presented. T$_{C}$ values have been estimated by taking the V$_{3}$ right onset. This is correct since T$_{C}$ and T$_{irr}$ merge at the limit of small fields. For $\lambda_{0}$ and $\xi_{0}$, well accepted average values presented in literature have been considered.

Measurements are performed fixing the coil current amplitude and increasing the temperature below T$_{C}$ after a zero field cooling. As an example, in figure 3a a set of measurements of V$_{3}$ versus T for the Nb sample are shown. By increasing the coil current, the applied magnetic field and consequently the total induction field on the sample surface increases, producing a broadening of the V$_{3}$ response. Let us  focus our attention on the curve having the highest intensity, obtained when the maximum induction field on the sample surface is about 0.8 mT: one can observe that as the temperature increases, V$_{3}$ remains zero until the first vortex penetrates the sample (left-onset). Then, it rises, reaching a maximum, and then decreases with a bell shape. The left onset  is $T(B^{\bot}_{C1})$, the temperature at which the total induction field at the sample surface (eq.~\ref{2}) is equal to the first critical field. As expected, when the total induction field is decreased, the left onset shifts monotonically towards $T_{C}(B=0)$ whereas the right onset remains quite unaffected.

The left-onset temperature is experimentally determined by looking at both the modulus and phase of the third harmonic signal temperature behavior (see arrows in figure 3b). In the amplitude versus temperature plot, the left/right onset temperature is defined as the first point whose amplitude exceeds the noise level floor (threshold criterion). In the specific case of figure 3b, this means that the sample signal third harmonic amplitude must exceed about 0.1 $\mu V$. The third harmonic phase has a random behavior (see the phase versus temperature plot) as it should be when the sample response is merely linear (in the Meissner state or in the flux flow regime and normal state) and no odd harmonics are present. Therefore, the left/right onset is defined as the temperature beyond which the third harmonic phase starts following a well defined pattern. The error is estimated by taking the difference between the $V_{3}$ onset temperature and the extrapolation of the linear portion of V$_{3}$ amplitude or phase vs temperature curve. This error never exceeds 0.5 \%. Therefore, one temperature scan gives one point in the B$^{\bot}_{C1}$ versus T curve. The same experiment is repeated with different field amplitudes for all samples under test, in order to evaluate the first critical field for a fairly large temperature range.

\section{DISCUSSION}

In figure 4 we show the measured B$^{\bot}_{C1}$ versus T curves for all samples under test. Experimental data have been compared to the expected theoretical values calculated by using the following equation\cite{BRANDTII,LYARDI,GUREVICHII,HARDY}:

\begin{eqnarray}\label{3}
B_{C1}\left( T \right) =\frac{\phi_{0}}{4 \pi \lambda^{2}\left(T\right) }\left[ ln\left( \kappa\right) + \alpha \left( \kappa \right) \right]
\end{eqnarray}

where $\kappa$ is the Ginzburg-Landau parameter, $\phi_{0}$ is the flux quantum, $\lambda$ is the London penetration depth and $\alpha \left( \kappa \right)$ is a small correction defined in ref. \onlinecite{BRANDTII}. $\alpha \left( \kappa \right) \sim 1/2$ for large $\kappa$ \cite{BRANDTII, HARDY, LYARDI}.

In figure 4a the experimental data (solid circles) and the calculated values for
$B^{\bot}_{C1}(t=T/T_{C})$ (continuous line) in the case of the Nb film are plotted. The calculation has been performed by using the normalized superfluid density of the BCS standard model\cite{MUHLS}. For $\kappa$, the temperature dependence reported in ref. \cite{KGINZBURG} has been considered. The experimental data are in good agreement with the theoretical predictions. Moreover, they well describe the tiny upward curvature observed in $B^{\bot}_{C1}(t)$ near the zero field critical temperature limit. As explained in detail below, the presence of this upward curvature could be ascribed to a number of reasons other than the temperature dependence of $\kappa$ itself \cite{KGINZBURG}. 

For the NbN film (figure 4b), the experimental data (solid circles) are in fair agreement with prediction, and measurements carried out in
ref.\cite{CHOCK} (open squares), however near $T_{C}$ only. The latter data have been
extracted from the temperature behavior of the upper critical field
taken by magneto-transport measurements on high quality films and
using the parameters collected in table 1. The continuous line
represents the theoretical prediction: the expected dependence
well agrees with the experimental data down to t=0.65. However it fails to justify the higher values measured at lower temperatures where $B^{\bot}_{C1}(t)$ is overestimated. The reason should be ascribed to the difficulty in using the assumption of an infinite slab, since the induction field on the sample edges cannot be neglected in comparison with the expected value of the first critical field. The screening is less effective than what expected from the model, which results in an overestimation of the current-field conversion factor. As reported in ref. \onlinecite{AURINO} for the specific case of our probing coil, the total induced magnetic field at the sample surface is completely negligible at the sample edges only if the ratio $r=r_{S}/r_{C}\gtrsim2.5$, where $r_{S}$ and $r_{C}$ are the sample and the external coil diameters respectively. By increasing the intensity of the current circulating in the probing coil, the total induction field spatial distribution extends over larger areas on the sample surface. When it exceeds sample dimensions, the total induction field at the edges can be greater than the expected first penetration field. This is actually what happens in the case of this film for temperature lower than 0.65, when the currents injected in the probing coil become greater than 35 mA. 

In the case of magnesium diboride c-axis oriented film (figure 4c), the measured $B^{\bot}_{C1}(t)$ coincides with $B^{c}_{C1}(t)$ because of our sample-coil configuration (see figure2b). We can note an excellent agreement between the experimental data (solid circles) and the theoretical prediction (continuous line) in the full temperature range by taking into account the two-band BCS model normalized superfluid density of ref. \onlinecite{GOLUBOV} in the clean limit. Our experimental results show that the first critical field in our sample is much higher than typical values for thin films \cite{ILONCA} (open circles) but lower than what reported for high quality single crystals
\cite{LYARDII,LYARDI} (open squares).

In the case of the YBCO film (figure 4d), experimental data are in fair agreement with the results obtained on single crystals shown in ref.\onlinecite{LIANG} and close to the values predicted by the two fluid model, again near the critical temperature only. For reduced temperatures lower than t=0.89, the measured first critical field is much larger than the values expected from the theoretical prediction. Again, as in the previous case of the NbN sample, this overestimation can be ascribed to the fact that the hypotesis of infnite slab fails when the current injected in the probing coil exceeds about 70 mA. The YBCO expected first critical field is higher than NbN one, that likley shifts the limit of validity of the infinite slab hypothesis to higer values of the probing coil current.

\section{CONCLUSIONS}

Some conclusive remarks can be made for all the investigated samples.
\textit{First} of all, the presence of an upward curvature is a common
feature present in all the experimental curves. This could be
explained by taking into account that
$B^{\bot}_{C1}(T)=B^{\bot}_{C2}(T)\cdot ln(\kappa)/(2\kappa^{2})$.
Thus, its presence in the upper critical
field should induce the same effect in the lower critical field. The upward cuvature has been indeed predicted for metals exhibiting anisotropic Fermi
surfaces \cite{KITA}, that is the case for quite all the systems
under study. However, particularly in multiband superconductors like MgB$_{2}$ the temperature dependencies of B$_{C1}$ and B$_{C2}$ can be very different and the Ginsburg-Landau parameter k depends on T. Besides that, in single-band s-wave superconductors, a downward curvature is expected based on both extensive measurements and calculations of B$_{C1}$ using the Eliashberg theory (see for example \onlinecite{CARBOTTE}). Likely, pinning may play a role here. \textit{Secondly}, the $B_{C1}$ values measured in our samples are in good agreement with data reported in literature for film samples of similar quality measured by other techniques. However, they are systematically lower than those reported for bulk samples. This is quite a general effect due to the reduced quality of the films in comparison with high quality bulk samples. 

In conclusion, we have presented a new and reliable method to measure the first critical field in superconducting type II materials under the form of thin film. This is done by looking at the third harmonic component of the response signal to an ac magnetic field
excitation. To validate this technique we have successfully
measured $B^{\bot}_{C1}(T)$ in low, medium and high critical temperature superconductors. 
The main factor limiting the technique as presented in this paper is the film size: only samples having an area larger than the coil dimensions at a given coil current can be measured. In particular, when sufficiently high currents are injected in the probing coil, and depending on the magnitude of the expected theoretical first critical field, the total induction field on the sample edges should be taken into account. This renders the hypothesis of the sample as an infinite slab questionable. As a consequence, an overestimation of the current-field conversion factor and of the measured first critical field will result.

\begin{acknowledgments}
We are grateful to P. Orgiani and R. Russo for providing us MgB$_{2}$ and Nb film samples. We would like to thank A. Gurevich, M. Polichetti, J. Halbritter, E. H. Brandt and Y. Mawatari for fruitful discussions and suggestions. We thank S. Zocco (Casale Bauer) for providing us the linear power amplifier SAMSON SX1200 and for fruitful discussions. We are indebted to V. De Luise for his valuable technical assistance. 
We acknowledge financial support of the Italian Foreign Affairs Ministry (MAE) - General Direction for the Cultural Promotion.\\
\\
\end{acknowledgments}

$^a$Author to whom correspondence should be addressed. Present address: Lamia, Department of Physics, University of Genoa, via Dodecaneso 33, 16100, Italy. E-mail: gianrico.lamura@lamia.infm.it.

$^b$Present address: LGEP-SUPELEC, 91190 Gif sur Yvette, France.

\newpage

\begin{table}
\label{table1} \caption{Summary of sample properties and parameters  for the evaluation of the first critical field. }

\begin{tabular}{|l|c|c|c|c|c|c|c|}
\hline
\textbf{sample}&{\textbf{thickness [nm]}}&{\textbf{substrate}}&{\textbf{T$_{C}$ [K]}}&{\textbf{$\lambda_{0}$[nm]}} &{\textbf{$\xi_{0}$[nm]}}\\
\hline \hline

\rm Nb&500&Si/Al$_{2}$O$_{3}$& 9.2 & 50 $^{[}$\cite{POOLE}$^{]}$& 40 $^{[}$\cite{POOLE}$^{]}$\\

\rm NbN&300&Al$_{2}$O$_{3}$&14.5&250 $^{[}$\cite{LAMURA,NBN,SENAPATI,CHOCK}$^{]}$&5 $^{[}$\cite{NBN,SENAPATI,CHOCK}$^{]}$\\

\rm MgB$_{2}$& 40 & SiC & 40.5 & 110 $^{[}$\cite{JIN,JINII}$^{]}$& 3.5 $^{[}$\cite{ORGIANI}$^{]}$ \\

\rm YBCO&700&Al$_{2}$O$_{3}$   & 87.9 & 150 $^{[}$\cite{POOLE}$^{]}$ & 1.5 $^{[}$\cite{SONIER}$^{]}$ \\

\hline 

\end{tabular}
\end{table}

\newpage

\textbf{Figure captions}

\textbf{Figure 1}: (color online) $V_{3}(T)$ behaviour for an a.c. magnetic field of amplitude $b_{0}=h_{0}/\mu_{0}$ for  a generic type II superconductor phase diagram. $V_{3}(T)$ is strictly equal to zero in the Meissner phase, it assumes finite values with a bell-shape behavior in the mixed state below the irreversibility line, and it comes back to zero in the flux flow and normal state regimes. The two vertical arrows refer to the left (T(B$_{C1}$)) and right (T(B$_{irr}$)) onset respectively. \\

\textbf{Figure 2}: a) sketch of the experimental set-up; b) details of the coil-sample configuration.\\

\textbf{Figure 3}: (color online) procedure for the extraction of the first critical field values from the temperature behavior of the third harmonic component (Nb sample): a) dependence of the $V_{3}$ left onset on the applied ac field by increasing the amplitude $I_{0}$ of the current injected in the probing coil. The arrow indicates the increase of $I_{0}$; b) details of the phase and modulus of the dotted curve shown in a). The arrow represents the third harmonic onset whereas the vertical line sets the method to determine the maximum amplitude error.\\

\textbf{Figure 4}: (color online) $B^{\bot}_{C1}$ measured values (solid \Large{$\circ$}\normalsize) as a function of the reduced temperature t=T/T$_{C}$ for the Nb (a), NbN (b), MgB$_{2}$ (c) and YBCO film (d). 
In (b), (\small{$\square$}\normalsize) data represent the experimental results reported in ref. \onlinecite{CHOCK}. In (c), (\small{$\square$}\normalsize) and (\Large{$\circ$}\normalsize) data  represent the experimental results reported in ref. \onlinecite{ILONCA} for thin films and in ref. \onlinecite{LYARDII,LYARDI} for single crystals respectively. In (d),  (\small{$\square$}\normalsize) data represent the experimental results reported in ref. \onlinecite{LIANG}. The continuous lines in each graph describe the theoretical predictions. \\
\\

\newpage

\Large\textbf{FIGURE 1}
\begin{figure}[hbtp!]
\includegraphics[scale=1.2]{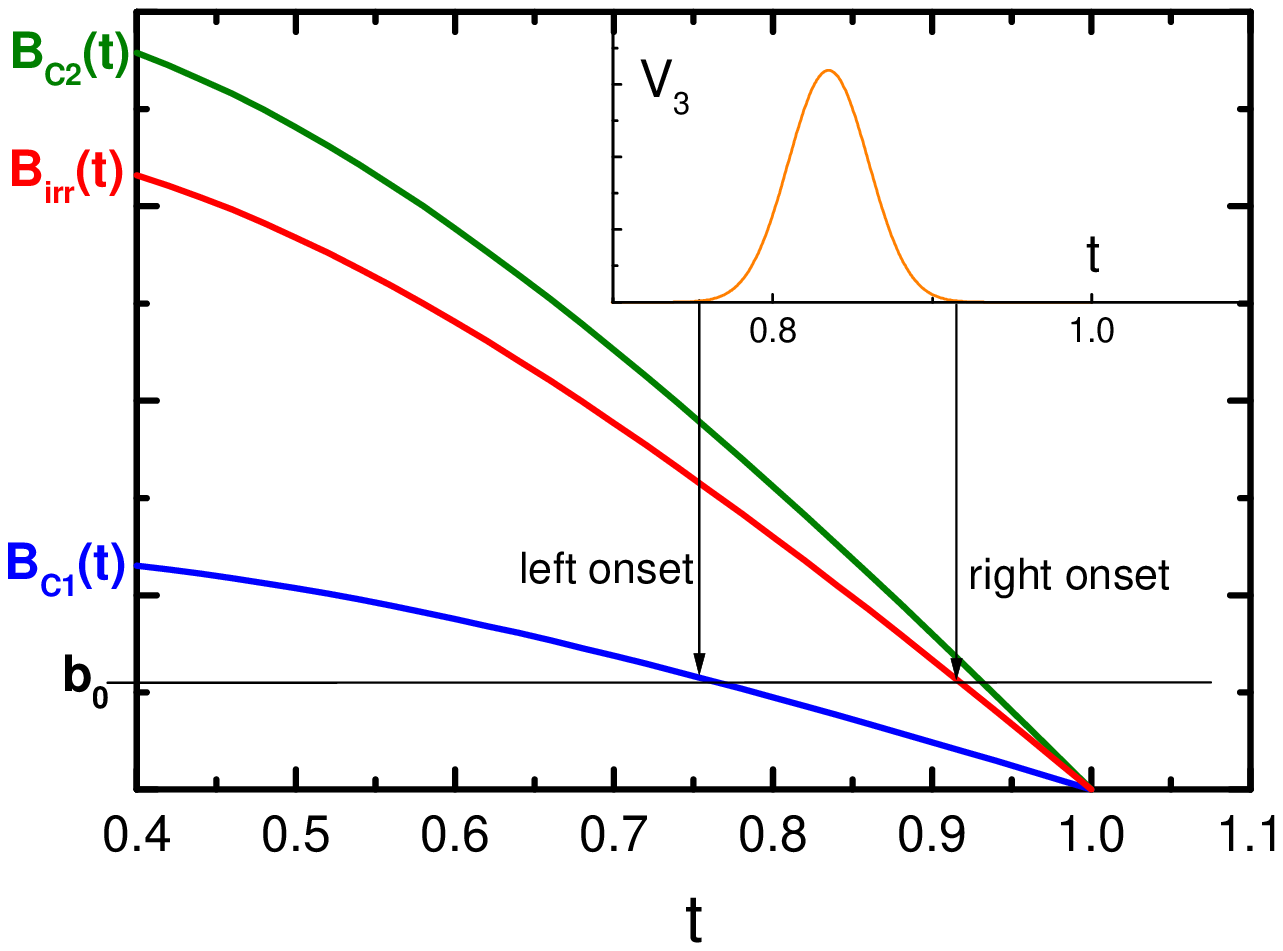}
\label{fig1}
\end{figure}

\newpage

\Large\textbf{FIGURE 2a}
\begin{figure}[hbtp!]
\includegraphics[scale=1.2]{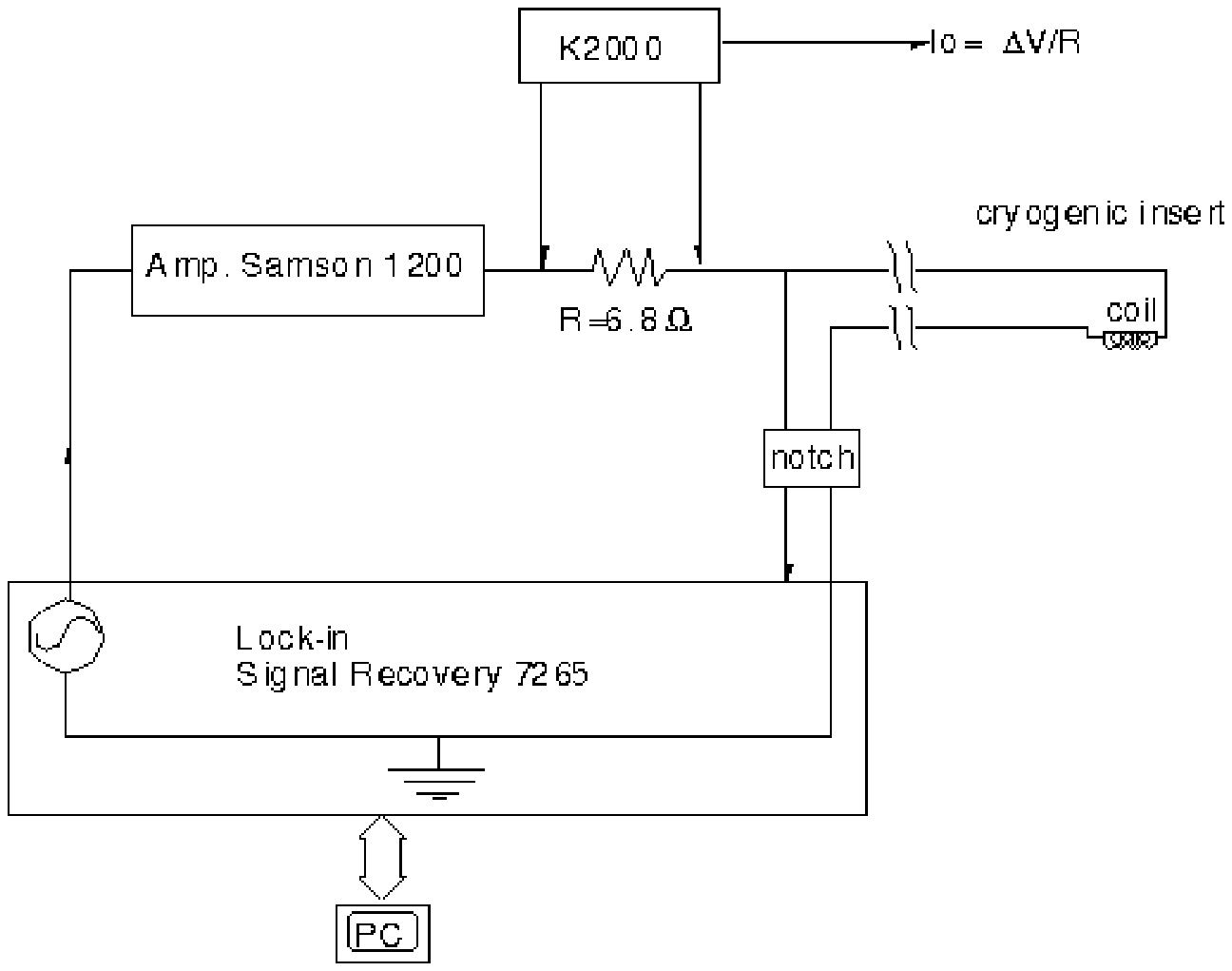}
\label{fig2}
\end{figure}

\newpage

\Large\textbf{FIGURE 2b}
\begin{figure}[hbtp!]
\includegraphics[scale=0.8]{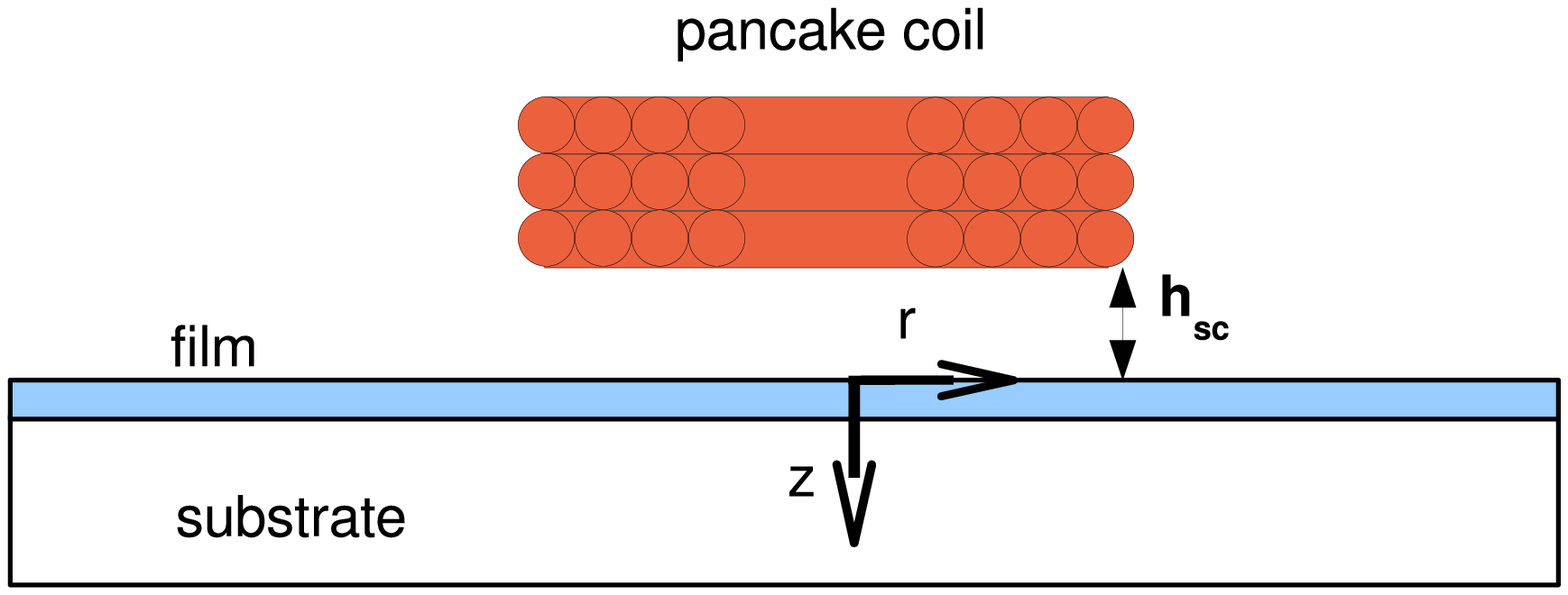}
\label{fig2}
\end{figure}

\newpage

\Large\textbf{FIGURE 3a}
\begin{figure}[h]
\includegraphics[scale=1.2]{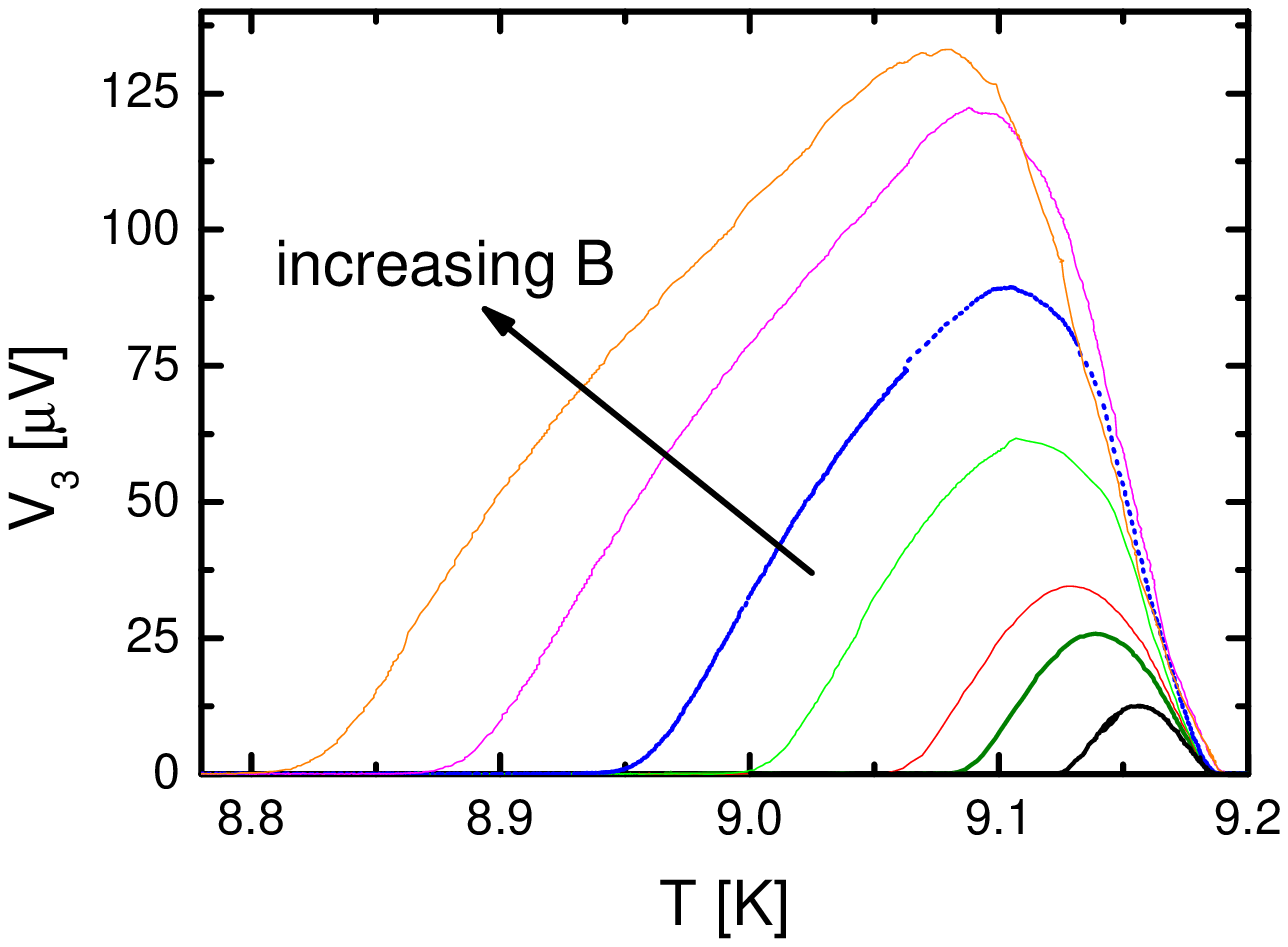}
\label{fig3a}
\end{figure}

\newpage

\Large\textbf{FIGURE 3b}
\begin{figure}[h]
\includegraphics[scale=1.2]{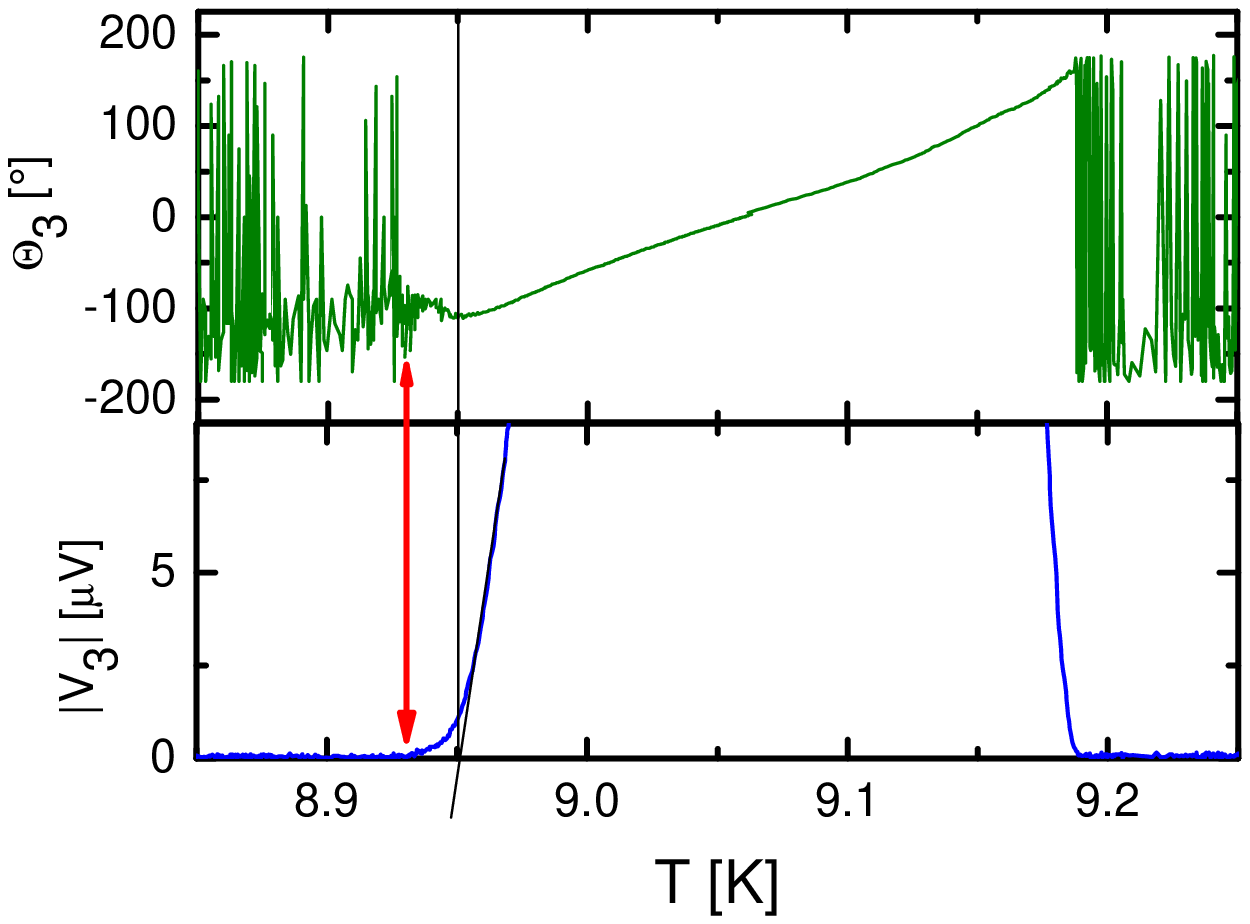}
\label{fig3b}
\end{figure}

\newpage

\Large\textbf{FIGURE 4a}
\begin{figure}[h]
\includegraphics[scale=1.2]{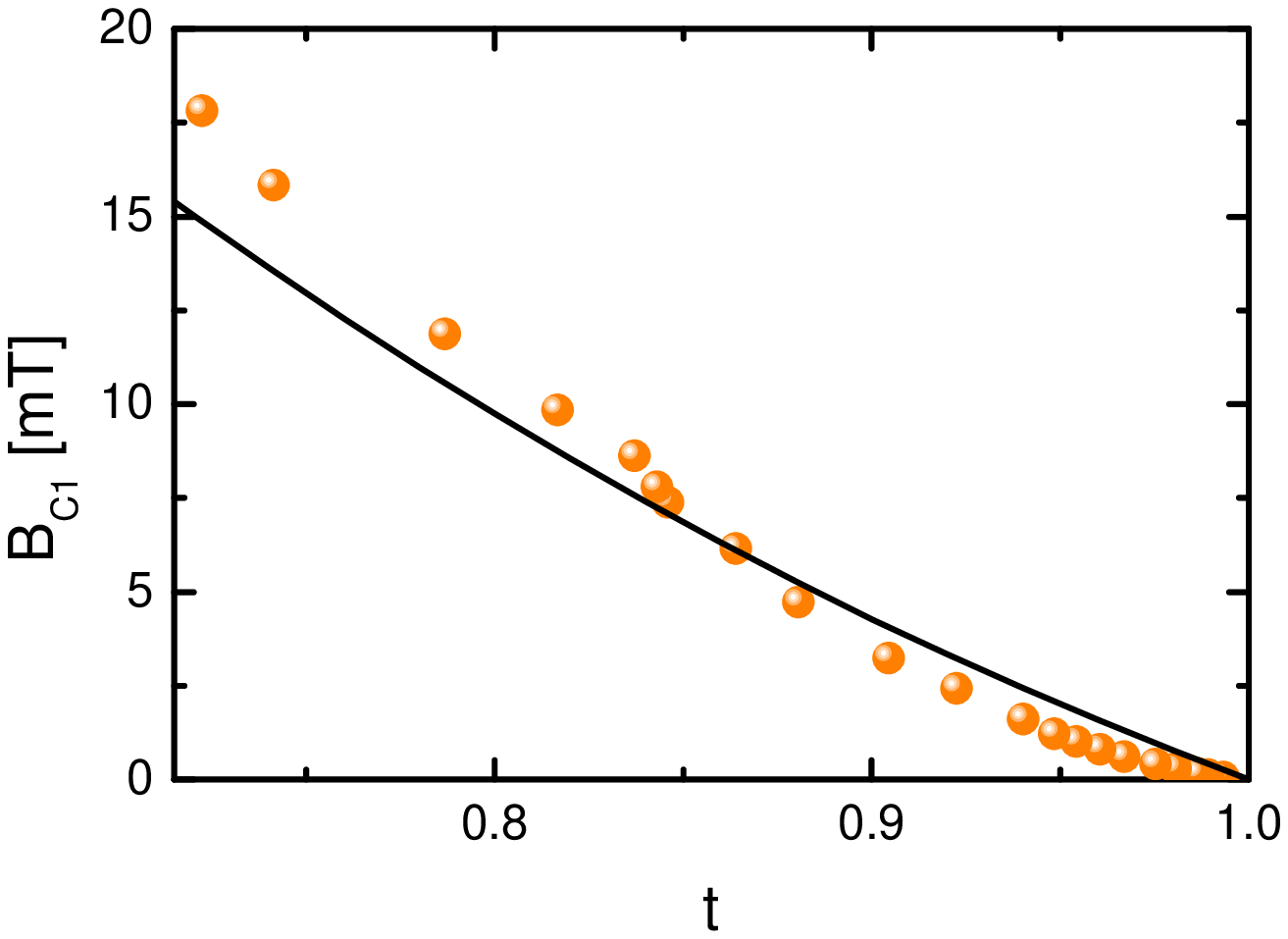}
\label{fig4a}
\end{figure}

\newpage

\Large\textbf{FIGURE 4b}
\begin{figure}[h]
\includegraphics[scale=1.2]{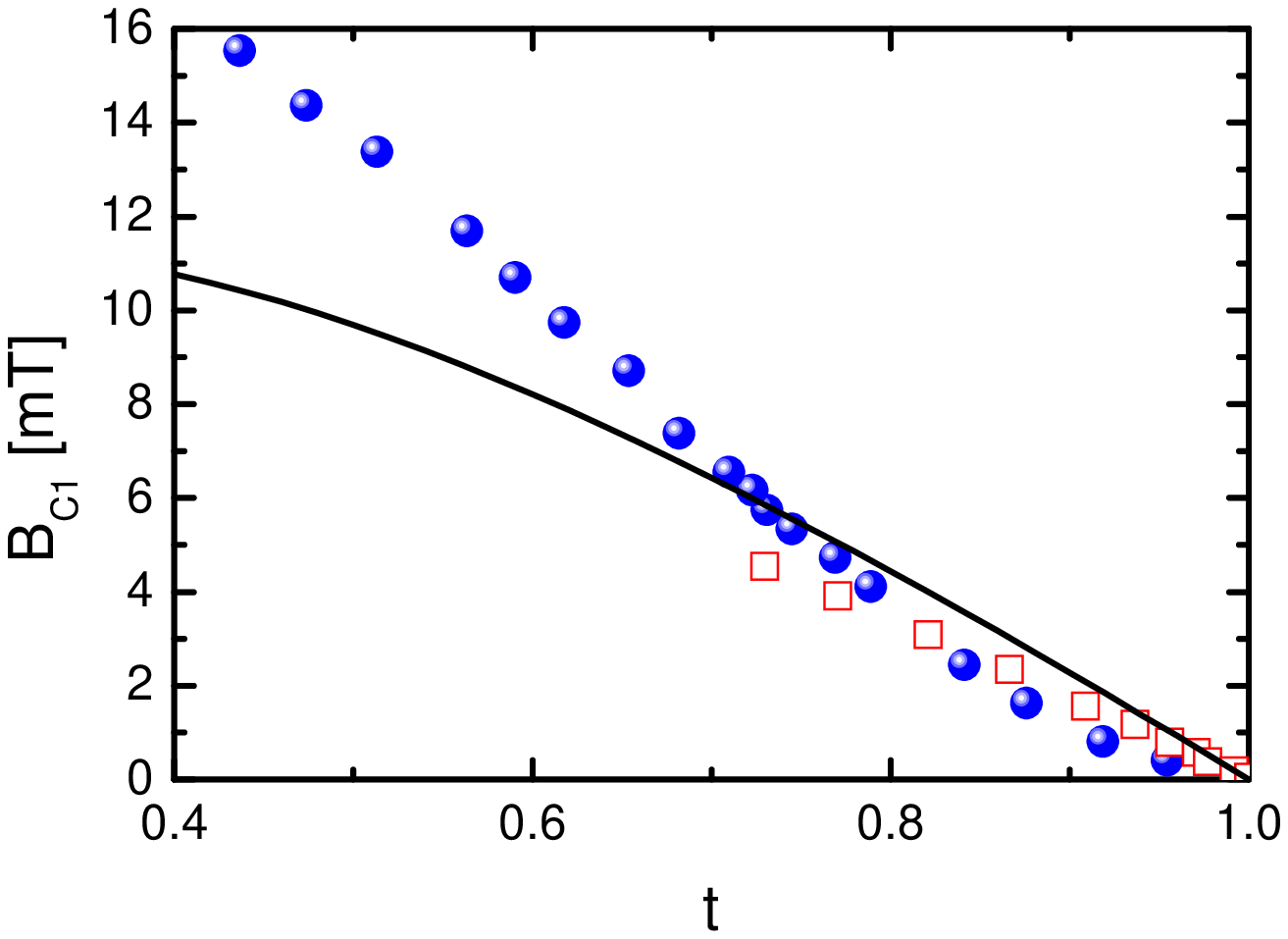}
\label{fig4b}
\end{figure}

\newpage

\Large\textbf{FIGURE 4c}
\begin{figure}[h]
\includegraphics[scale=1.2]{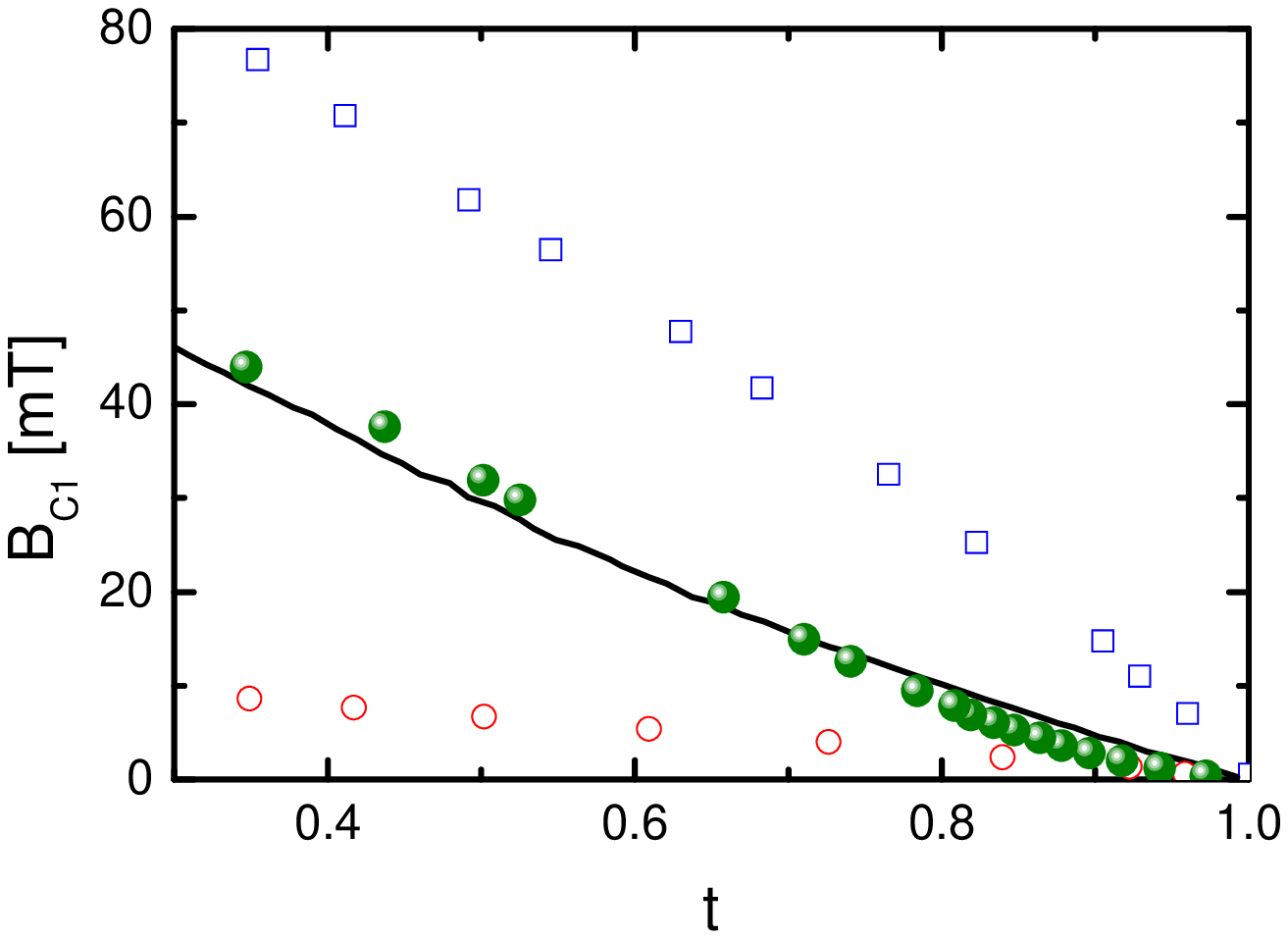}
\label{fig4c}
\end{figure}

\newpage

\Large\textbf{FIGURE 4d}
\begin{figure}[h]
\includegraphics[scale=1.2]{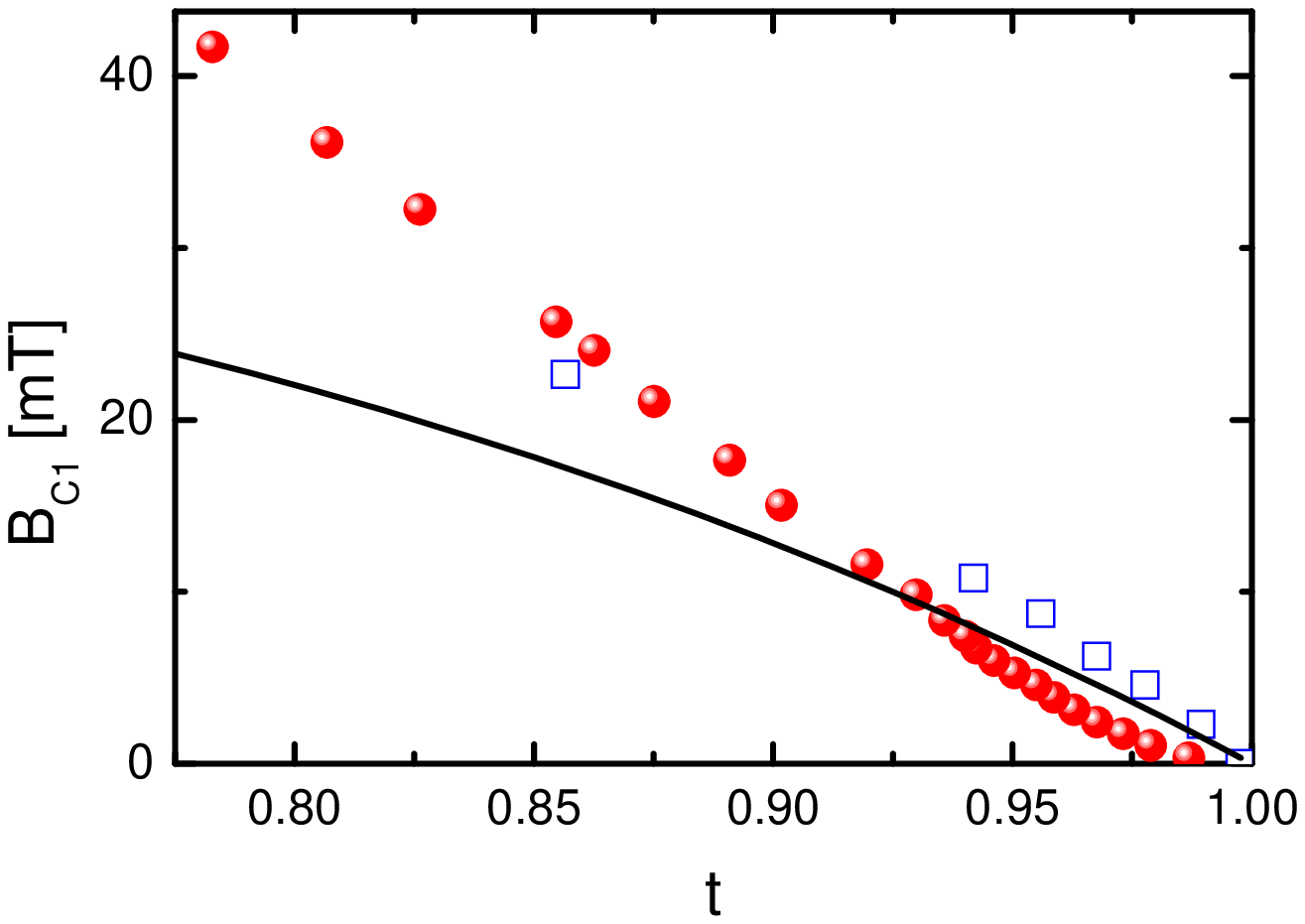}
\label{fig4d}
\end{figure}

\end{document}